\documentclass[12pt]{article}
\usepackage{graphicx}

\title{Longitudinal Losses Due to Breathing Mode
 Excitation in Radiofrequency Linear Accelerators
}
\author{Paul J. Channell\footnote{pchannell@lanl.gov}\\
MS H808\\
Los Alamos National Laboratory\\
Los Alamos, NM 87545}
\date{}
\def\ie{i.e.\ }
\def\be{\begin{equation}}
\def\ee{\end{equation}}
\def\ni{\noindent}

\begin{document}

\ni {\Large LA-UR-10-07086}\vspace{0.25in}

\ni {\it {\small Approved for public release:}}
\vspace{0.01in}

\ni {\it {\small distribution is unlimited.}}
\vspace{0.6in}

\hspace{1.0in}{\it {\small Title:Longitudinal Losses Due to Breathing Mode

\hspace{1.3in} Excitation
in Radiofrequency Linear Accelerators}}
\vspace{0.5in}

\hspace{1.0in}{\it {\small Author(s):Paul J. Channell}}
\vspace{0.5in}

\hspace{1in}{\it {\small Intended for:}}
\vspace{0.5in}

\hspace{-0.5in} \includegraphics{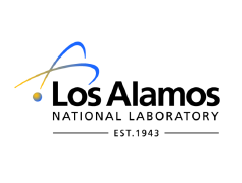}
\vspace{0.1in}

\ni {\tiny Los Alamos National Laboratory, an affirmative action/equal opportunity
employer, is operated by the Los Alamos National Security, LLC
for the National Nuclear Security Administration for the U.S. Department of
Energy under contract DE-AC52-06NA25396. 
By acceptance of this article, the publisher recognizes that the U.S. Government retains a nonexclusive, royalty-free
license to publish or reproduce the published form of this contribution,
or to allow others to do so, for U.S. Government purposes. Los Alamos 
National Laboratory requests that the publisher identify this article as
work performed under the auspices of the U.S. Department of
Energy. Los Alamos National Laboratory strongly supports academic freedom and
a researcher's right to publish; as an institution, however, the Laboratory
does not endorse the viewpoint of a publication or guarantee its technical correctness.}
\vspace{0.01in}

\ni {\tiny Form 836 (7/06)}

\newpage

\maketitle

\begin{abstract}
Transverse breathing mode oscillations in a particle beam can couple
energy into longitudinal oscillations in a bunch of finite length
and cause significant losses. We develop a model that illustrates
this effect and explore the dependence on mismatch size, space-charge
tune depression, longitudinal focusing strength, bunch length, and
RF bucket length. 
\end{abstract}

\newpage

\section{Introduction}

In high average power radiofrequency (RF) linacs the problem of losses can
be acute, as they tend to activate the structure. Depending on the
average beam power, loss fractions less than one part in a million may
be required. Previously people, \cite{gluckstern}, \cite{wangler}, \cite{wangler2},
 have examined the role of transverse
breathing mode oscillations on transverse losses and concluded that
they could drive particles out to several times the original beam size.
Of course, in modern superconducting RF accelerators with strong focusing,
the wall radius can be many times the beam size so these oscillations
may not lead to unacceptable losses.

However, these breathing mode oscillations can also couple to the longitudinal
oscillations in a finite length bunch and cause particles to move outside
the RF bucket. In that case, particles will eventually have an energy so
far away from the design energy that they will be lost transversely. Note
that typically the length of the RF bucket is only a few times the length
of the particle bunch (for protons or ions) so a detailed investigation is required.

In this paper we examine the effect of transverse breathing mode oscillations
on the longitudinal oscillations of particles and show that in some
circumstances the resulting losses can be very significant.

\section{Model}

In this article we will investigate beam loss in the longitudinal
direction caused by a type of deterministic diffusion in a  finite length 
mismatched beam. A mismatch causes envelope oscillations that are 
approximately described by the 
equation

\begin{equation}
{d^2X\over dt^2} +X - {\eta ^2\over X^3}-{I\over X}=0,
\end{equation}

\noindent where $X$ is the envelope amplitude, $\eta$ is
the tune depression parameter, and $I\equiv 1-\eta ^2$.
Note that if $X=1$ and ${dX\over dt}=0$ then the breathing
mode has zero amplitude, \ie the envelope doesn't oscillate.

The transverse motion of a single particle in the resulting
oscillating hard-edge beam has been described by Gluckstern, Wangler
and others, \cite{gluckstern} \cite{wangler}, for one transverse
degree of freedom and is given by the equation

\begin{equation}
0={d^2x\over dt^2} + x -\left\{ 
\begin{array}{l l}
   & {I x\over X^2} \quad \mbox{if $x < X$}\\
   & {I\over x} \quad \mbox{if $x > X$}\\ \end{array} \right. 
\end{equation}

\noindent The discontinuity in derivative
at $x=X$ is awkward, so we replace this equation by

\begin{equation}
0={d^2x\over dt^2} + x -{I x\over x^2+X^2},
\end{equation}

\noindent a continuous version which has the same behavior as the orginal equation.
The transverse phase space diagram for this equation is the standard
`peanut' shape shown in figure 1.

\begin{figure}[h]
\includegraphics[scale=0.55]{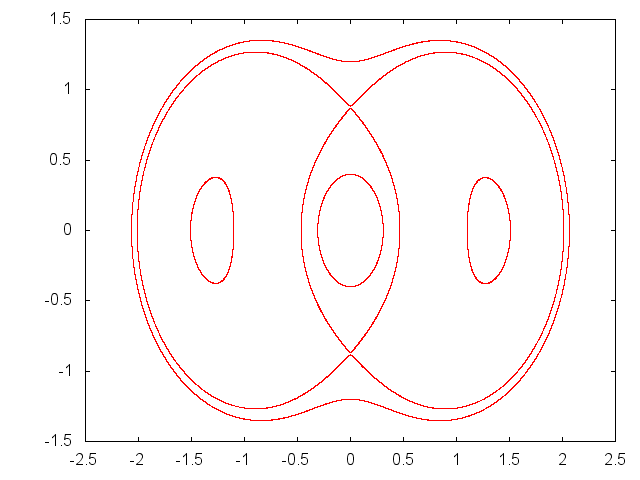}
\caption{Five transverse trajectories of the uncoupled system
illustrating the usual `peanut' phase space.}
\end{figure}

We now want to incorporate the longitudinal motion of of particles
in a finite length bunch. Without coupling we take the longitudinal
motion to be described by

\begin{equation}
0={d^2z\over dt^2} + {\omega_z^2b\over 2\pi }\sin ({2\pi z\over b}),
\end{equation}

\noindent i.e. a simple pendulum with small amplitude angular oscillation frequency $\omega_z$,
and an effective length of $b$, the `bucket' length. In a real RF linac the bucket
shape is more complicated but this model should capture the effect of 
nonlinear longitudinal oscillations and
a finite
length region of longitudinal stability.
Note that these equations for the particle motion can be derived from the Hamiltonian

\begin{equation}
H={p_x^2\over 2}+{x^2\over 2}+{p_z^2\over 2}-{\omega_z^2 b^2\over 4 \pi ^2}\cos ({2\pi z\over b})-{I\over 2}\ln (x^2+X^2).
\end{equation}

In order to have a simple model of the coupling of longitudinal and transverse motion,
we let 

\begin{equation}
I=I_0 e^{-z^2\over L^2},
\end{equation}

\noindent where $L$ is a bunch length. Note that in a real finite length bunch the envelope
oscillation would not retain the simple form we have assumed here; thus our
model is overly simplified but should be indicative of the type of physics that can occur.
The Hamiltonian now becomes

\begin{equation}
H={p_x^2\over 2}+{x^2\over 2}+{p_z^2\over 2}-{\omega_z^2 b^2\over 4 \pi ^2}\cos ({2\pi z\over b})-{I_0 e^{-z^2\over L^2}\over 2}\ln (x^2+X^2).
\end{equation}

\noindent The equations of motion that result from this coupled Hamiltonian are

\begin{equation}
{dx\over dt}=p_x,
\end{equation}

\begin{equation}
{dp_x\over dt} = -x +{I_0 e^{-z^2\over L^2} x\over x^2+X^2},
\end{equation}

\begin{equation}
{dz\over dt}=p_z,
\end{equation}

\noindent and

\begin{equation}
{dp_z\over dt} = -{\omega_z^2b\over 2\pi}\sin ({2\pi z\over b}) +{I_0  z\over L^2}\ln (x^2+X^2)e^{-z^2\over L^2}.
\end{equation}

\noindent Note that as $L \rightarrow \infty$ these equations reduce to the
uncoupled equations.

 The driving mechanism for the particle instability and loss we observe in this model is the transverse envelope
oscillation and it's modulation due to the longitudinal change in space charge strength along the bunch. 
Another term could come from coupling transverse space charge oscillations to
longitudinal space charge oscillations which would also drive particles longitudinally. 
A simple model of this effect does not yet exist, so we won't include this term in this paper.

\subsection{Model Parameters}

The model we use has a number of parameters that can be varied. 
The transverse angular betatron frequency is 
$\omega _\beta=1$ and does not vary. The longitudinal angular oscillation frequency,
$\omega _z$, is a measure of the strength of the longitudinal focusing.
The bunch length is measured by $L$; the actual bunch length ($95 \% $ included) is about
$4L$ if the bunch is Gaussian. The bucket length, $b$, depends on the accelerating phase
and may be half of the RF wavelength (no acceleration), but is
typically less than a quarter
of the RF wavelength. Note that the bucket ranges from $-b/2$ to $b/2$
in $z$. 
The tune depression parameter is $\eta$ and ranges 
from $1$ (no space charge) down to small values
for high space charge. Finally there is the initial value of
the envelope oscillation, $X_0$, which is $1$ for no oscillation.
(We always take the initial ${dX\over dt} =0$.) We take various
values of the initial $X_0$ ranging from $0.9$ down to $0.45$; the
envelope oscillation is nonlinear but is such that the maximum
$X$ is approximately $2-X_0$; the oscillation is very roughly
symmetrical about $X=1$. The period of the envelope oscillation
depends on the amplitude of the oscillation, but is near $4$ for the
range we consider.

\subsection{Reason for instability}

To understand the reason the transverse envelope oscillation can drive longitudinal
oscillations consider figure $2$ where we plot in three dimensions the trajectories
of two points starting at the same transverse phase space location but with different
longitudinal oscillation amplitudes. 
What is shown is a stroboscopic plot taken
at exactly period intervals of the envelope oscillation, a cross
section in mathematical terminology. The picture shown has
no longitudinal-transverse coupling. Each trajectory is a product of an ellipse
in the longitudinal phase space (we show only one longitudinal phase space
variable) with an unstable trajectory of the `peanut' in
the transverse phase space. Note that the transverse unstable ellipse 
(point when projected only onto the transverse plane) at the
cusp of the shown trajectories is actually an ellipse crossed into a circle since
time is periodic and doesn't appear in the stroboscopic slice; thus it is a 
two-dimensional torus. The trajectories shown (which are three dimensional when time
is included) lie on what Arnold, \cite{arnold}, calls `whiskers' of the whiskered tori
which are the two dimensional unstable points crossed ellipses crossed periodic time.
Note that in this case the whiskers don't look much like real whiskers so
the name is misleading.
In the pictured uncoupled case these whiskers don't intersect and no instability
can occur. However, when coupling is added to the model, the whiskers from two
different longitudinal oscillation amplitudes do generically intersect and
connect initial conditions with small longitudinal oscillation amplitudes to
trajectories with large longitudinal oscillation amplitudes, leading to instability
and particle loss. Thus, the energy of the transverse envelope oscillation can be
pumped into the longitudinal oscillations of particles leading to longitudinal
losses. Of course it is important to determine the magnitudes of the losses for
various parameters and determine the most important variables to be controlled. 
Determining the lossses analytically would be very difficult so we will use simulation
of the above model to get an indication of loss rates.

\begin{figure}
 \begin{center}
\includegraphics[scale=0.3]{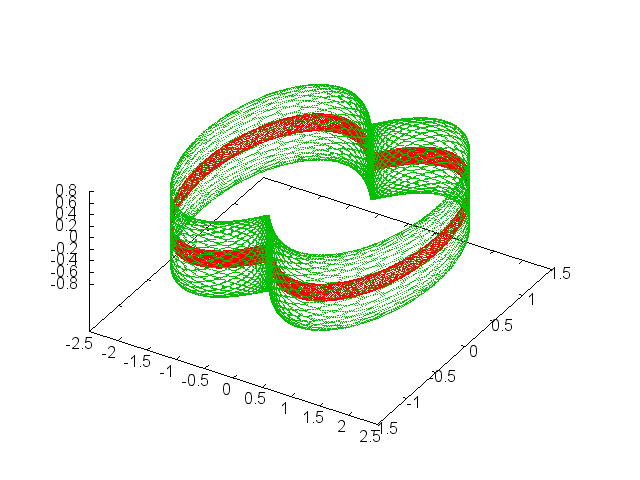}
\caption{Two trajectories (one red, one green) starting at the same transverse location but
with different longitudinal energies. These lie on `whiskers' of the
unstable whiskered tori.}
 \end{center}
\end{figure}

\section{Simulation}

All simulations were done using a fourth order symplectic integrator,
\cite{yoshida}. For each set of parameters and envelope initial
condition the envelope period was first determined using a very
tiny time step, \ie to double precision accuracy. Particle trajectories
were then printed out at multiples of this envelope period so that
stroboscopic pictures could be made. The envelope oscillation was
computed numerically at each time step along with the trajectories
of the particles.

In all the simulations reported here the particles were initialized
with very small longitudinal momentum. In longitudinal position the
particles were chosen to be uniformly random in the interval 
$[-0.67L,0.67L]$ where $L$ is the bunch length parameter; 
note that we are 
 not using the same distribution for the test (simulation) particles that
we assume for the space-charge model. In the transverse direction the
initial conditions were chosen to be uniformly random in both $x$ and
$p_x$ in the interval $[-0.335,0.335]$ where the unstable transverse
point is at about $[0,0.87]$, (it varies depending on parameters).

 Integrations were carried out with $2000$ steps per envelope period
and the integration was carried out for $100$ envelope periods. The number
of particles in the simulations was $4000$.
Particles were counted as lost when the absolute magnitude of their
longitudinal position exceeded $b/2$.

 Simulations were done varying
$X_0$ over the values $(0.45,0.6,0.75,0.9)$,
 $\eta$ over the values $(0.35,0.5,0.65,0.8)$,
$\omega_z^2$ over the values $(0.01, 0.11, 0.21, 0.31)$,
the bucket length, $b$, over the values $(10,15,20,25)$,
and the bunch length parameter, $L$, over the values $(2,3,4,5,6)$.
 A total of  $704$ simulations were done, namely those that had bunch lengths
($4L$) less than the bucket length ($b$). We report the results of these $704$
simulations.

\section{Results}  

Of the $704$ cases, $158$ led to particle losses ranging from $1$ particle (out
of $4000$) to $100\%$ of the particles. More than $10\%$ losses occured in $101$ cases,
and greater than $1\%$ losses occured in a further $27$ cases.
A phase space plot of a case with only $1\% $ particles lost is shown in figure $3$.

\begin{figure}[t]
 \begin{center}

\hspace{-0.3in} \includegraphics[scale=0.4]{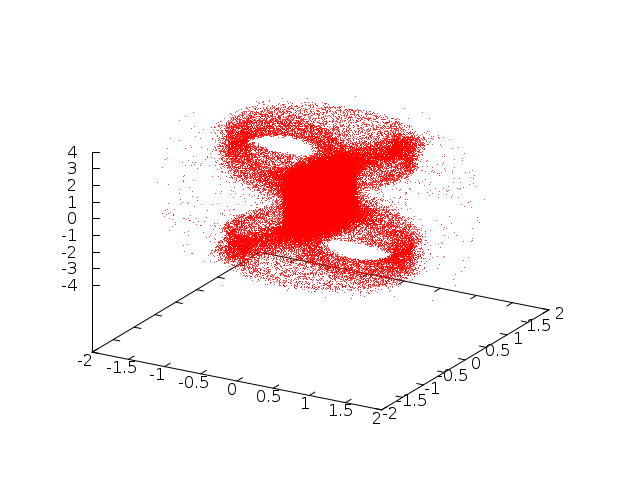}

\caption{
Stroboscopic phase space plot of a case with only $1\%$ of the particles lost. 
The three dimensional projection  ($x,p_x,z$) of $4000$ particles at
$100$ envelope periods are shown; though  
not all points are shown as
the lost particles move out to larger longitudinal positions.
}
 \end{center}
\end{figure}

\begin{table}
\begin{center}
\begin{tabular}{c | c}
Variable & Correlation coefficient \\
\hline
$X_0$ & -0.2885519  \\
$\eta$ &  -0.1515029 \\
$\omega_z^2$ & -0.4810484 \\  
$b$ & -0.1672322 \\
$L$  & -0.0471679  \\
\end{tabular}
\caption{Correlation coefficients of losses with model variables.}
\end{center}
\end{table}

The correlation coefficients of the particle
losses with the variables are shown in table $1$.
As can be seen, all the variables affect the particle losses, though the dependence
on $L$ is quite weak. The directions of the correlations
are what one would expect, \ie one has lower losses with a) smaller mismatch, b)
less space charge (less current), c) stronger longitudinal focusing, d) longer
bucket lengths, and e) shorter bunches.

In more detail, note that all values of mismatch, even small ones ($X_0=0.9$), can lead to
particle losses. Any particle loss in our simulations, a part in $4000$, is greater than
most high power linacs can allow. Also, particle losses can occur for all values of the
space charge, including very low space charge, $\eta=0.9$, though they tend to be worse
for higher space charge. The longitudinal focusing strength in our simulations ranged
from moderate ($\omega_z^2=0.01$) to very strong, ($\omega_z^2=0.31$ which is nearly
$1/3$
the transverse focusing strength) and particle losses occured for 
all cases.
The strongest dependence on the model parameters was on the
longitudinal focusing strength. No one parameter can guarantee no losses, though there
are only a few cases with loss for the largest longitudinal focusing.

In table $2$ we give the parameters and fractional losses for the thirty  cases with 
largest losses. Note that all values of mismatch, space charge, bucket length, and
bunch length occur. However, only the smallest longitudinal focusing frequency occurs
for these high loss cases. Unfortunately, this lowest longitudinal focusing strength
is also the case most typical of RF linacs. Note that by assuming small mismatch
and small space charge it is possible for cases with this lowest longitudinal focusing strength
to have no losses.

\begin{table}
\begin{center}
\begin{tabular}{|c | c | c | c | c | c|}
\hline
$X_0$ & $\eta$ & $\omega_2^2$ & $b$ & $L$  & Loss \\
\hline
    0.45 &   0.35 &   0.01 &   10. &   2. &   1.  \\
    0.45 &   0.35 &   0.01 &   15. &   3. &   1.  \\
    0.45 &   0.50 &   0.01 &   10. &   2. &   1.  \\
    0.45 &   0.50 &   0.01 &   15. &   3. &   1.  \\
    0.45 &   0.65 &   0.01 &   10. &   2. &   1.  \\
    0.45 &   0.80 &   0.01 &   10. &   2. &   1.  \\
    0.45 &   0.35 &   0.01 &   15. &   2. &   0.9995  \\
    0.60 &   0.50 &   0.01 &   10. &   2. &   0.99875  \\
    0.45 &   0.50 &   0.01 &   15. &   2. &   0.9965  \\
    0.45 &   0.65 &   0.01 &   15. &   3. &   0.99625  \\
    0.60 &   0.65 &   0.01 &   10. &   2. &   0.991  \\
    0.45 &   0.35 &   0.01 &   20. &   3. &   0.985  \\
    0.45 &   0.65 &   0.01 &   15. &   2. &   0.98275  \\
    0.45 &   0.50 &   0.01 &   20. &   4. &   0.979  \\
    0.45 &   0.50 &   0.01 &   20. &   3. &   0.96125  \\
    0.45 &   0.35 &   0.01 &   20. &   2. &   0.95925  \\
    0.45 &   0.65 &   0.01 &   20. &   4. &   0.93275  \\
    0.60 &   0.50 &   0.01 &   15. &   2. &   0.9215  \\
    0.45 &   0.50 &   0.01 &   25. &   4. &   0.914  \\
    0.45 &   0.35 &   0.01 &   25. &   3. &   0.90075  \\
    0.60 &   0.35 &   0.01 &   10. &   2. &   0.8845  \\
    0.60 &   0.80 &   0.01 &   10. &   2. &   0.882  \\
    0.45 &   0.65 &   0.01 &   20. &   3. &   0.881  \\
    0.45 &   0.35 &   0.01 &   25. &   2. &   0.859  \\
    0.60 &   0.35 &   0.01 &   15. &   2. &   0.841  \\
    0.45 &   0.35 &   0.01 &   20. &   4. &   0.8195  \\
    0.45 &   0.35 &   0.01 &   25. &   4. &   0.813  \\
    0.45 &   0.35 &   0.01 &   25. &   6. &   0.80075  \\
    0.45 &   0.65 &   0.01 &   25. &   6. &   0.8005  \\
    0.45 &   0.50 &   0.01 &   25. &   5. &   0.788  \\
\hline
\end{tabular}
\caption{Thirty cases with largest fractional losses.
}
\end{center}
\end{table}

If we look at the projections of the fractional losses on the
various two dimensional subspaces of the variables we get the three dimensional
figures shown.
In figures $5$b, $7$a, $8$a, and $8$b one notes the very weak dependence on the bunch
length found in the correlation coefficients.
In figures $4$b, $6$a, $7$b, and  $8$a one notes the strong dependence on longitudinal
focusing strength also found in the correlation coefficients. 
The moderate dependence of losses on mismatch can be seen in figures $4$a, $4$b, $5$a, and $5$b.
The weak dependence of losses on space charge can be seen in figures
$4$a, $6$a, $6$b, and $7$a. The  dependence of losses on bucket length
can be seen in figures $5$a, $6$b, $7$b, and $8$b; it seems to be significant only
for the shortest buckets.

\subsection{An extended model}

At the suggestion of Yuri Batygin we have extended the previous model to include
RF nonlinearity; in particular we studied the model Hamiltonian

\begin{equation}
H={p_x^2\over 2}+{x^2\over 2}+{p_z^2\over 2}-{\omega_z^2 b^2\over 4 \pi ^2}\cos ({2\pi z\over b})-{I\over 2}\ln (x^2+X^2)
+ {b\alpha x^2\over 2\pi}\sin({2\pi z\over b}),
\end{equation}

\ni where $\alpha$ controls the magnitude of the RF nonlinearity.
Setting the bunch length parameter, $L = 3.0$, we performed $768$ simulations with
$$\alpha = 0.003,0.006,0.009,0.012;$$ 

\ni the largest value gives a nonlinear betatron frequency  shift at the ends of the bunch of
about $\pm 9\%$. Note that the additional term is antisymmetric about the bunch center and
this antisymmetry can be seen in the phase space plot in figure $4$.

\begin{figure}[t]
 \begin{center}

\hspace{-0.3in} \includegraphics[scale=0.4]{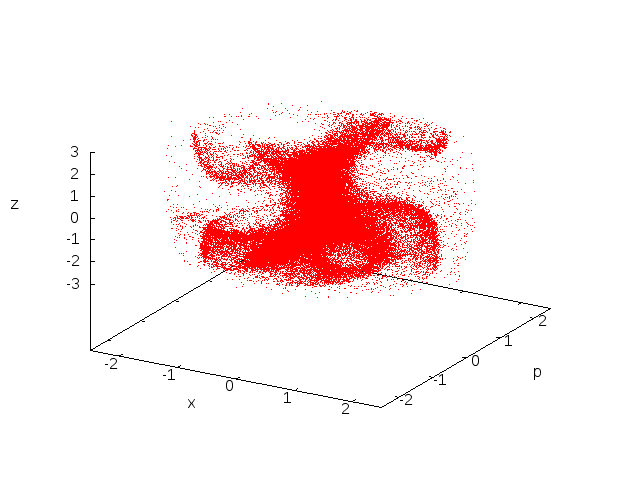}

\caption{
Stroboscopic phase space plot of a case with only $1\%$ of the particles lost. 
The three dimensional projection  ($x,p_x,z$) of $4000$ particles at
$100$ envelope periods are shown; though  
not all points are shown as
the lost particles move out to larger longitudinal positions.
Note the asymmetry in $z$ evident in the plot.
}
 \end{center}
\end{figure}

The results of these simulations with the extended model are essentially the same as those
from the previous model. In particular the correlation coefficient of the losses with the
$\alpha $ parameter is $0.0035$, \ie is consistent with $0$. The projection graphs for
this case are essentially the same as figures $5 - 9$ (when the variables coincide)
so we will not show them again.

\section{Conclusions}

No single parameter guarantees no losses in our simulations, though smaller
mismatch and larger longitudinal focusing have the strongest effect in 
reducing losses. Less space charge and longer bucket lengths have significant,
though smaller effects on losses and bunch length is relatively unimportant.
The nonlinearity of the RF seems to have very little effect on the losses.

\vspace{-0.3in}
\begin{figure}
 \begin{center}
\includegraphics[scale=0.5]{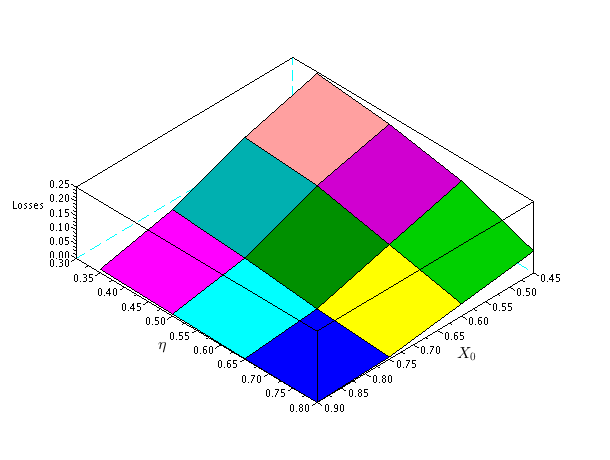}
(a)
\includegraphics[scale=0.5]{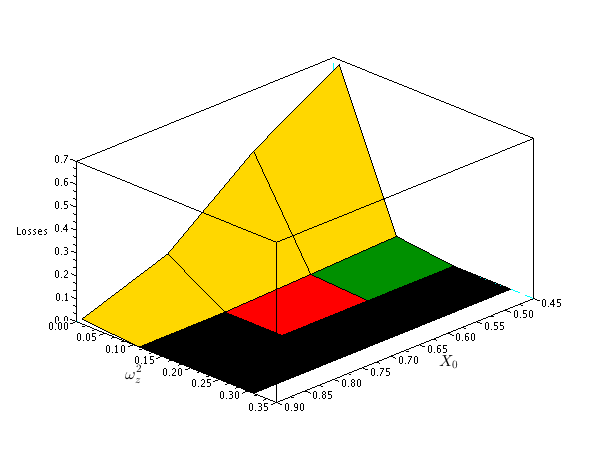}
(b)
\caption{Projected losses as functions of
a) $X_0$ and $\eta$, b) $X_0$ and $\omega_z^2$. Smaller losses
occur for smaller mismatch ($X_0$ closer to $1$), smaller
space charge ($\eta$ closer to $1$), and larger longitudinal focusing,
larger $\omega_z^2$, with this dependence rather strong.
}
 \end{center}
\end{figure}

\vspace{-0.3in}
\begin{figure}
 \begin{center}

\includegraphics[scale=0.5]{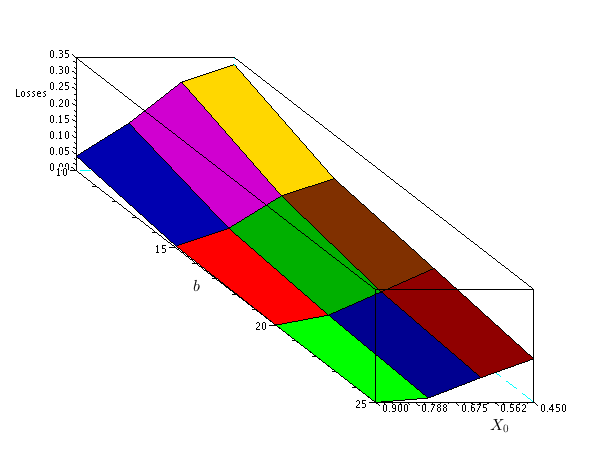}
(a)
\includegraphics[scale=0.5]{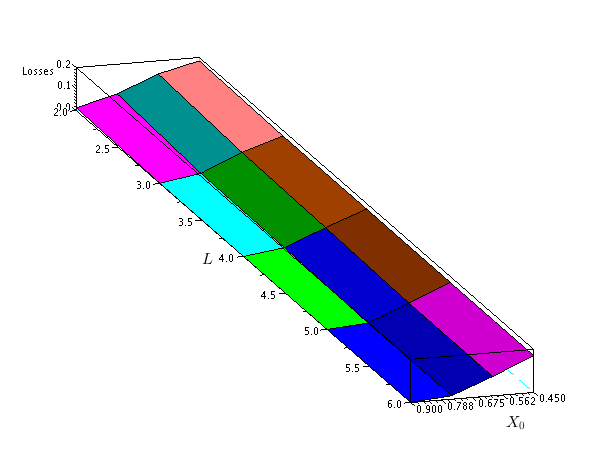}
(b)
\caption{Projected losses as functions of
a) $X_0$ and $b$, and b) $X_0$ and $L$.
Smaller losses occur for larger bucket lengths, $b$,
and shorter bunches, $L$, though the latter dependence is quite weak.
The dependence on bucket length is strongest for short buckets.
}
 \end{center}
\end{figure}

\vspace{-0.3in}
\begin{figure}
 \begin{center}
\includegraphics[scale=0.5]{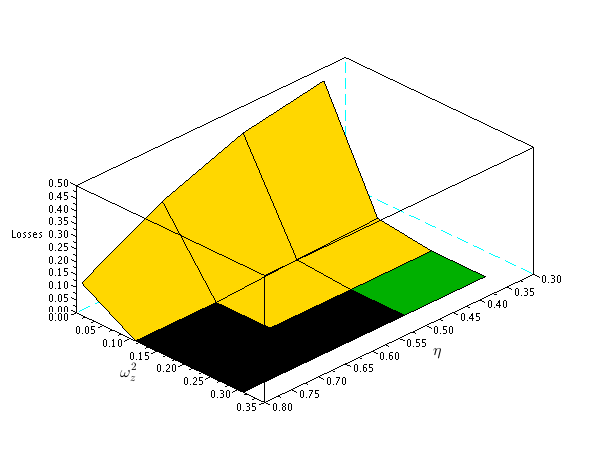}
(a)
\includegraphics[scale=0.5]{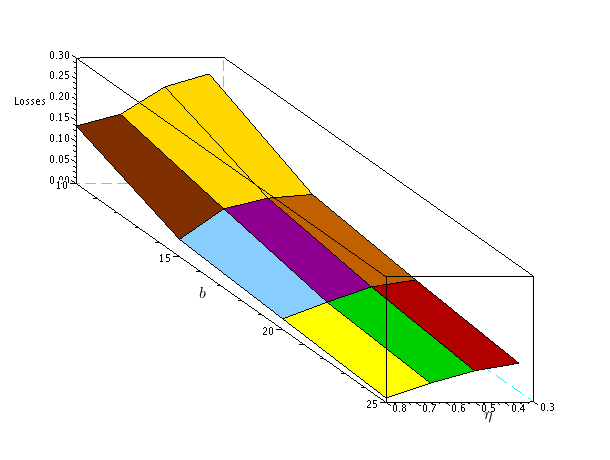}
(b)
\caption{Projected losses as functions of
a) $\eta$ and $\omega_z^2$, b) $\eta$ and $b$.
Losses are smaller for less space charge, ($\eta$ larger),
stronger longitudinal focusing, and larger bucket lengths.
}
 \end{center}
\end{figure}

\vspace{-0.3in}
\begin{figure}
 \begin{center}
\includegraphics[scale=0.5]{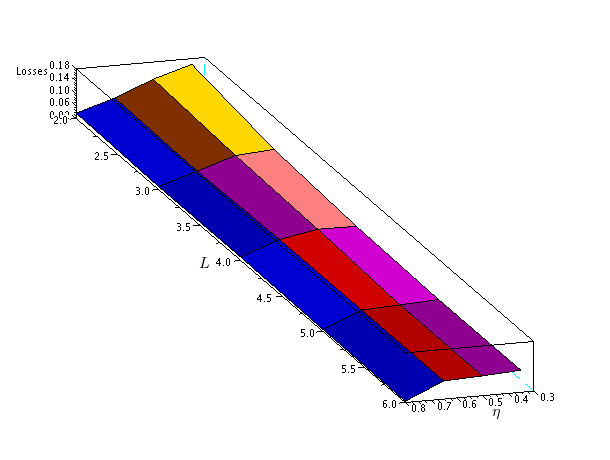}
(a)
\includegraphics[scale=0.5]{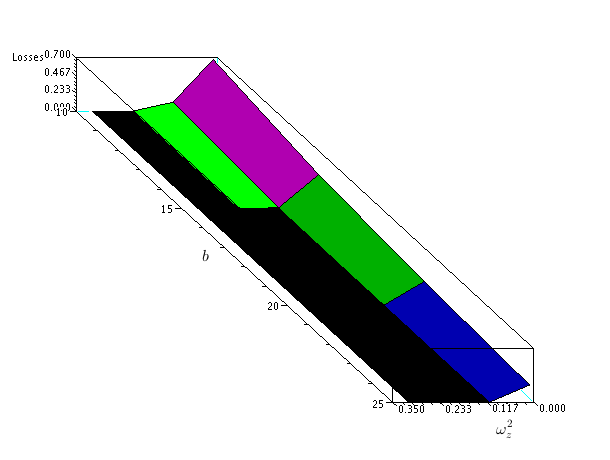}
(b)
\caption{Projected losses as functions of
a) $\eta$ and $L$, and b) $\omega_z^2$ and $b$.
The losses depend weakly on bunch length and on bucket length
 and more strongly on longitudinal
focusing strength.
}
 \end{center}
\end{figure}

\vspace{-0.3in}
\begin{figure}
 \begin{center}
\includegraphics[scale=0.5]{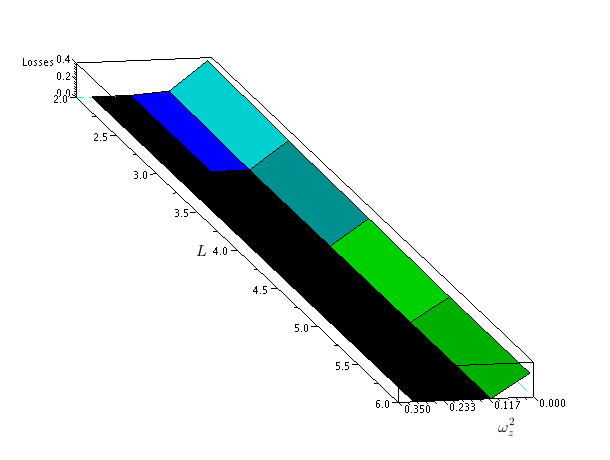}
(a)
\includegraphics[scale=0.5]{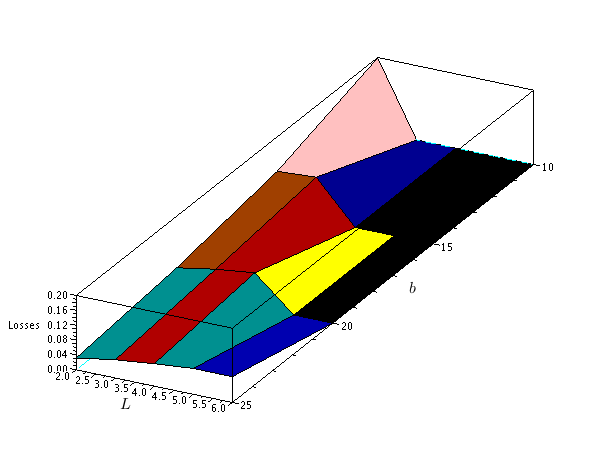}
(b)
\caption{Projected losses as functions of
a) $\omega_z^2$ and $L$, b) $b$ and $L$.
the losses depend most strongly on $\omega_z^2$.
}
 \end{center}
\end{figure}

\vspace{1.8in}

\ni {\bf Acknowledgments}
Conversations with and suggestions from Yuri Batygin and Tom Wangler are greatly appreciated.

\end{document}